\documentclass[aps, prc, reprint, superscriptaddress]{revtex4-1}
\usepackage{amsmath}
\usepackage{amsfonts}
\usepackage{amssymb}

\usepackage{graphicx}
\usepackage[caption=false]{subfig}

\usepackage[colorlinks]{hyperref}
\hypersetup{linkcolor=black, citecolor=black, filecolor=black, urlcolor=black}
\usepackage[all]{hypcap}

\usepackage[utf8]{inputenc}

\newcommand{\IOPP}{Key Laboratory of Quark \& Lepton Physics (MOE) and Institute of Particle Physics, \\
Central China Normal University, Wuhan 430079, China}

\begin{document}

\title{Effects of resonance weak decays and hadronic re-scattering on the proton number fluctuations in Au + Au collisions at $\sqrt{s_\mathrm{NN}} = 5$~GeV from JAM model}
\author{Yu Zhang \footnotemark[1]}
\author{Shu He \footnotemark[2]} 
\author{Hui Liu}
\author{Zhenzhen Yang}
\affiliation{\IOPP}
\author{Xiaofeng Luo}
\email{xfluo@mail.ccnu.edu.cn}
\affiliation{\IOPP}
\footnotetext[1]{You can add acknowledgements here.}
\footnotetext[2]{You can add acknowledgements here.}

\begin{abstract}
Proton number fluctuation is sensitive observable to search for the QCD critical point in heavy-ion collisions. In this paper, we studied rapidity acceptance dependence of the proton cumulants and correlation functions in most central Au+Au collisions at $\sqrt{s_\mathrm{NN}} = 5$ GeV from a microscopic hadronic transport model (JAM).   At mid-rapidity, we found the effects of resonance weak decays and hadronic re-scattering on the proton cumulants and correlation functions are small, but those effects get larger when further increasing the rapidity acceptance.  On the other hand, we found the baryon number conservation is a dominant background effect on the rapidity acceptance dependence of proton number fluctuations. It leads to a strong suppression of cumulants and cumulant ratios, as well as the negative proton correlation functions. We also studied those two effects on the energy dependence of cumulant ratios of net-proton distributions in most central Au+Au collisions at $\sqrt{s_\mathrm{NN}} = 5-200$~GeV from JAM model.  This work can serve as a non-critical baseline for future QCD critical point search in heavy-ion collisions at high baryon density region.
\end{abstract}

\maketitle

\section{Introduction}
Exploring the QCD phase structure is one of the main goals of heavy-ion collision experiments. It can be displayed in the QCD phase diagram, which is a two dimensional $T-\mu_\mathrm{B}$ plane. Lattice QCD calculations confirmed that the transition from Quark-Gluon Plasma (QGP) to hadronic phase at the zero baryon chemical potential ($\mu_\mathrm{B}=0$) is smooth crossover~\cite{Aoki:2006we}. QCD based models predict a first order phase transition at large $\mu_\mathrm{B}$~\cite{Ejiri:2008xt}. If both of the crossover and first order transition are true, there must be an end point of the first order phase transition boundary, which is so called QCD critical point. The experimental and/or theoretical confirmation of the QCD critical point would be a landmark in exploring the QCD phase structure.

Fluctuations of conserved charges, such as net-baryon ($B$), net-charge ($Q$) and net-strangeness ($S$), are sensitive probes to the QCD critical point and phase transition in heavy-ion collisions~\cite{Ejiri:2005wq, Stephanov:2008qz, Stephanov:2011pb, Schaefer:2011ex, Asakawa:2009aj}. These observables have been extensively studied experimentally~\cite{Aggarwal:2010wy, Adamczyk:2014fia, Adamczyk:2013dal}
and theoretically~\cite{Karsch:2010ck, BraunMunzinger:2011dn, Gavai:2010zn, Chen:2014ufa, Chen:2015dra, Kitazawa:2011wh, Kitazawa:2012at, Friman:2011pf, Mukherjee:2015swa, Mukherjee:2016kyu, Vovchenko:2015pya, Jiang:2015hri, Nahrgang:2014fza, Morita:2014fda, Bazavov:2012vg, Borsanyi:2013hza, Alba:2014eba, Fan:2016ovc,Fan:2019gkf,FanWenKai:2019imq,Li:2018ygx,Ye:2018vbc}. 
\begin{figure}[!htb] \label{fig:KV}
	\includegraphics[width=2.8in]{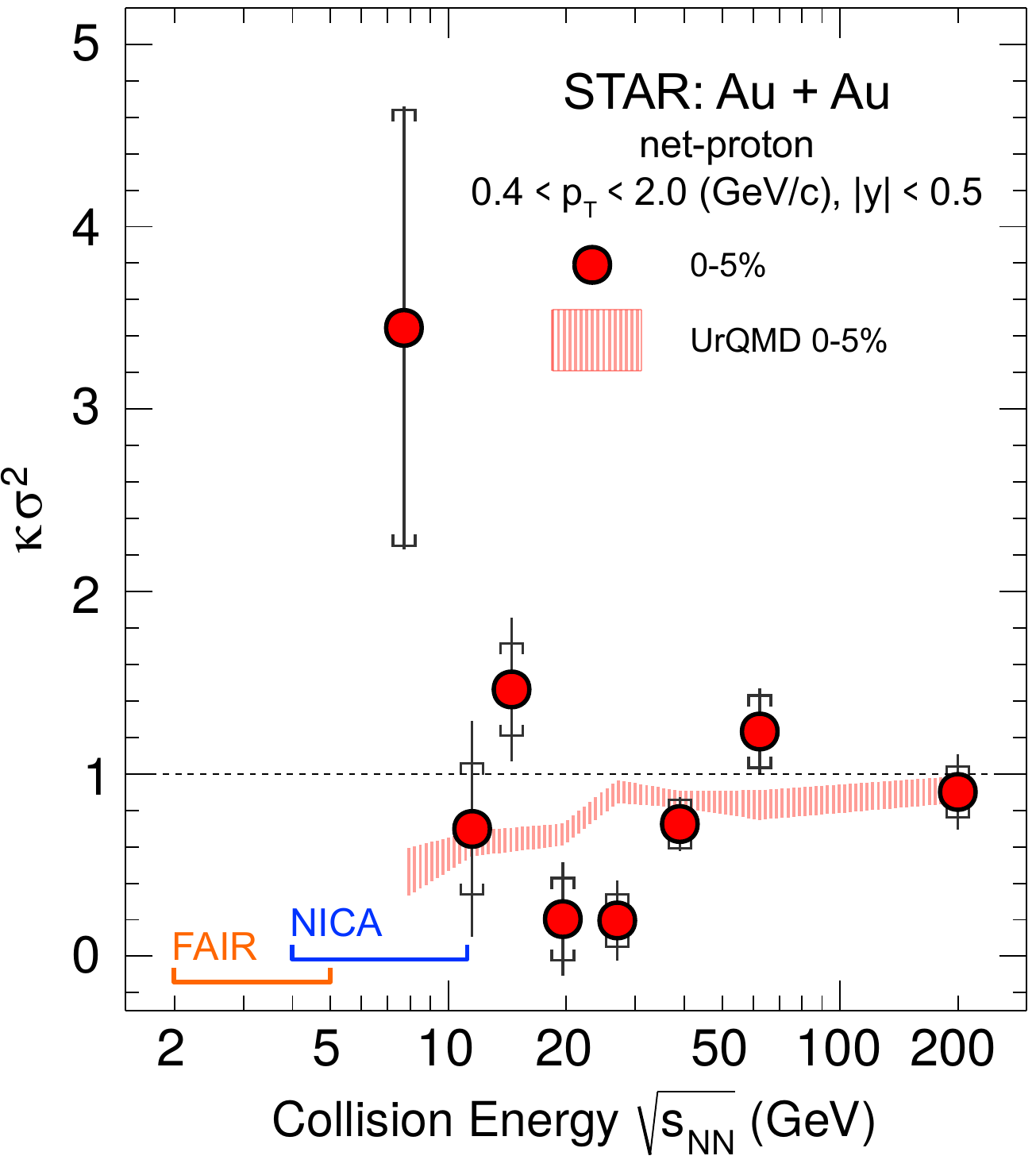}
	\caption{Energy dependence of fourth order cumulants ratio ($\kappa \sigma^{2}$) of net-proton multiplicity distributions from STAR experiment~\cite{Adam:2020unf}. The energy coverage of the FAIR and NICA heavy-ion programs are marked as orange and blue caps in the plot, respectively.}
\end{figure}
In the year 2010 to 2014, RHIC has finished the first phase of beam energy scan (BES) and took data of Au+Au collisions at $\sqrt{s_\mathrm{NN}}=7.7, 11.5, 14.5, 19.6, 27, 39, 62.4, 200$~GeV. With those experimental data, STAR experiment has measured the higher order fluctuations of net-proton, net-charge and net-kaon multiplicity distributions~\cite{Adamczyk:2013dal, Adamczyk:2014fia, Luo:2015doi,Adamczyk:2017wsl,Luo:2017faz}. As shown in Fig.~\ref{fig:KV}, one of the most striking observation is the non-monotonic energy dependence of the fourth order cumulants ratio($\kappa \sigma^2$) of the net-proton/proton number fluctuations in most central (0-5\%) Au+Au collisions~\cite{Luo:2017faz}.  It was observed that the fourth order net-proton fluctuation is close to unity above 39 GeV but deviates significantly below unity at 19.6 and 27 GeV, then becomes above unity at lower energies. This non-monotonic structure is predicted by models assuming the existence of critical point~\cite{Stephanov:2011zz,Bluhm:2016byc,Bluhm:2016trm,Herold:2016uvv,Mukherjee:2016nhb,Nahrgang:2016eou,Vovchenko:2015pya}. This may suggest that the created system skims close by the critical point, and received positive and/or negative contributions from critical fluctuations.  On the other hand, the enhancement of the  $\kappa\sigma^2$ at low energies cannot be described by the UrQMD model~\cite{Xu:2016qjd,Zhou:2017jfk}, which does not contain the physics of critical point. For second and third order net-proton cumulant ratios ($C_3/C_2$ and $C_2/C_1$), they show deviations below from the Poisson expectations~\cite{ Luo:2015doi}, and are dominated by the contributions from baryon number conservation (BNC).  

To extract the signature of critical fluctuations, it is crucial to understand the background contributions for proton number fluctuations in heavy-ion collisions. Some of the background contributions, such as baryon number conservations~\cite{Bzdak:2012an}, acceptance/efficiency corrections~\cite{Bzdak:2012ab,Luo:2014rea,Luo:2018ofd}, light nuclei formation~\cite{Feckova:2015qza}, initial volume fluctuations, auto-correlation, and the effects of centrality selections~\cite{Luo:2013bmi,Zhou:2018fxx,Sugiura:2019toh,Chatterjee:2019fey},  have been studied before. However,  these studied background effects have difficulties to describe the $\kappa\sigma^2 \gg 1$ at low energies. A phenomenological model study shows that the large increase of net-proton $ \kappa\sigma^2$ above unity at $\sqrt{s_\mathrm{NN}}=7.7$ GeV can be explained as the formation of multi-protons clusters~\cite{Bzdak:2016sxg,Bzdak:2017ltv}. One may note that, in Refs.~\cite{Bzdak:2016sxg,Bzdak:2017ltv}, the $n^{th}$ order cumulants and correlation functions are denoted as $\kappa_n$ and $C_n$, respectively, which is opposite to what we used in the current paper. 

 In this work, we performed detailed studies for the effects of resonance weak decays and hadronic re-scattering on the proton number fluctuations in most central Au+Au collisions $\sqrt{s_\mathrm{NN}}=5$ GeV with JAM model. The energy is chosen, because it will be covered by the future FAIR/CBM and NICA/MPD experiments. The resonance weak decays and hadronic re-scattering can be turned on or off in the JAM model. For hadronic re-scattering, we studied two effects : one is the effects of meson-baryon (MB) and meson-meson (MM) interactions, and the other is the hadronic elastic scattering. Finally, we show the energy dependence of net-proton cumulants ratios in most central Au+Au collisions at $\sqrt{s_\mathrm{NN}}=5-200$ GeV from JAM model. 

This paper is organized as follows, we first introduce the fluctuation observables: cumulants and correlation functions in \autoref{sec:cumulants_correlations}. Then we introduce the JAM model in \autoref{sec:effect_in_jam}. In \autoref{sec:effect_to_fluctuation}, we present the results of proton cumulants and correlation functions, and discuss the effects of resonance weak decays and hadronic re-scattering. Finally, we give a summary.
\begin{figure}\label{fig:IPP}
\hspace{-0.5cm}
	\includegraphics[width=2.8in]{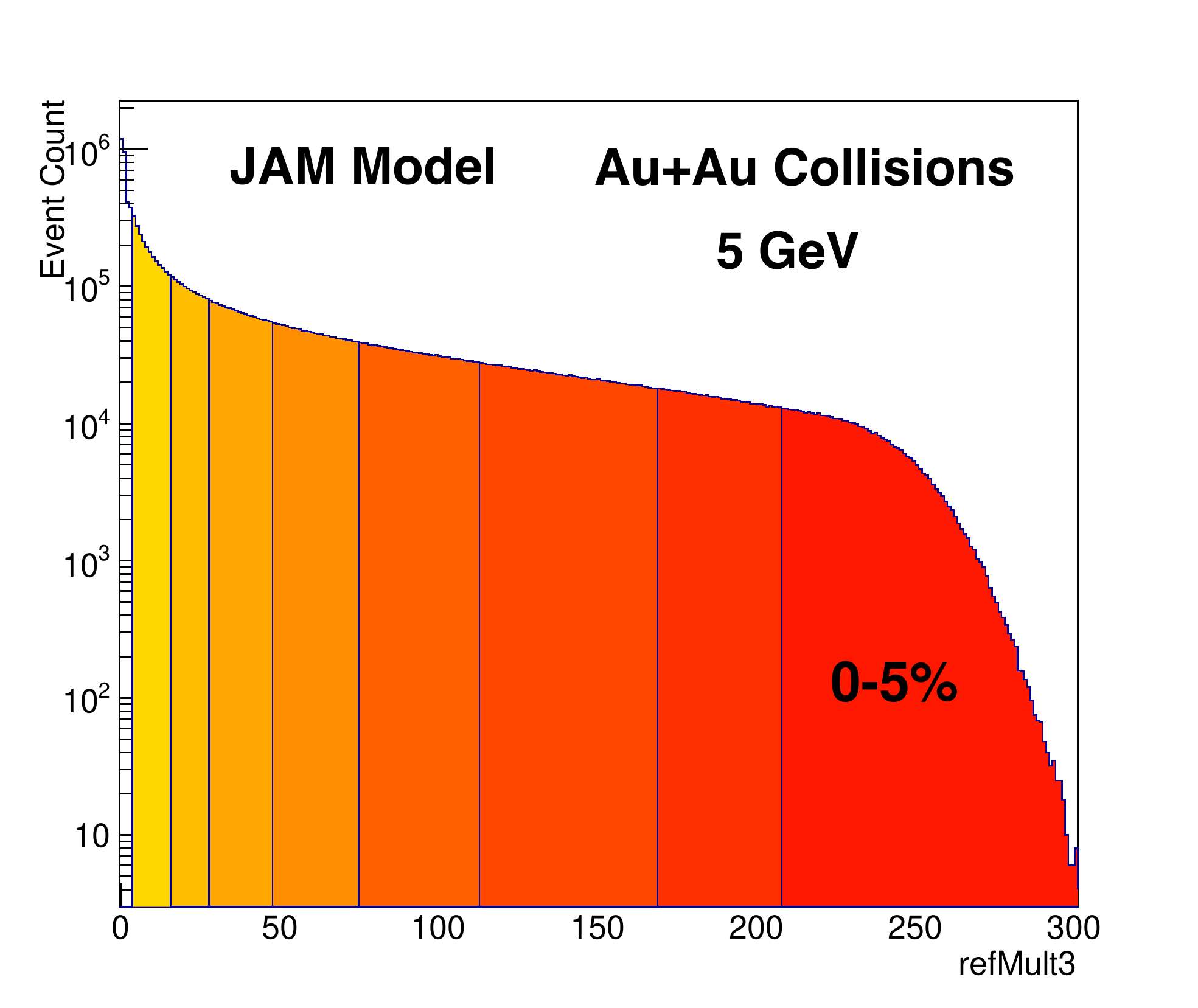}
	\vspace{-0.5cm}
	\caption{The refMult3 distributions in Au+Au collisions at $\sqrt{s_\mathrm{NN}} = 5$~GeV from JAM model. It is defined as number of charged (anti)pions and (anti)kaons within $|\eta|<1$. Protons and anti-protons are excluded from this definition to avoid auto-correlation effects. Events from top 5\% centrality class is used in this analysis.}
\end{figure}

\section{Cumulants and Correlation Functions}\label{sec:cumulants_correlations}
To characterize the multiplicity fluctuations, one can measure the cumulants of the particle multiplicity distributions.  The various order cumulants are calculated from moments as:
\begin{equation}\label{eq:moment_to_cumulant}
\begin{aligned}
    C_1 = &\langle N \rangle \\
    C_2 = &\langle N^2 \rangle - \langle N \rangle^2 \\
    C_3 = &2\langle N \rangle^3 - 3\langle N\rangle \langle N^2\rangle + \langle N^3 \rangle \\
    C_4 = &-6\langle N\rangle^4 + 12\langle N\rangle^2 \langle N^2\rangle - 3\langle N^2\rangle^2 \\
          &-4\langle N\rangle \langle N^3\rangle + \langle N^4 \rangle 
\end{aligned}
\end{equation}
, where the $\langle N^n \rangle$ is the $n^{th}$ order moment of particle number distributions. The $n^{th}$ order cumulant $C_n$ is connected to the susceptibilities $\chi_n$ of system as~\cite{Cheng:2008zh}
\begin{equation}\label{eq:cumulants_and_correlation}
    C_n = VT^3 \chi_n
\end{equation}
To cancel out the volume $V$, the ratios of different order of cumulants are usually constructed as experimental observables:%
\begin{equation}\label{eq:cumulant_ratio}
\begin{aligned}
      S\sigma  &= \frac{C_3}{C_2}=\frac{\chi_3}{\chi_2},                &\kappa\sigma^2 &= \frac{C_4}{C_2}=\frac{\chi_4}{\chi_2}
\end{aligned}
\end{equation}
, where $S$ and $\kappa$ are skewness and kurtosis of the multiplicity distributions respectively.

The collision centralities are defined by using charged pions and kaons at mid-rapidity ($|\eta|<$1), which is so called refMult3. In our study, as shown in Fig.~\ref{fig:IPP}, only top 5\% centrality is used in the calculations. The centrality bin width correction (CBWC)~\cite{Luo:2011ts, Luo:2013bmi} is also applied to suppress volume fluctuations in a wide centrality bin. In CBWC method, as shown in Eq. (\ref{eq:CBWC}), the cumulants are calculated for event ensemble in each refMult3 bin ($i$) and are taken an average with number of events ($n_i$) as the weights for each bin. 
\begin{equation}\label{eq:CBWC}
\begin{aligned}
      C^{}_{r}&= \frac{\sum n_i C^{i}_{r}}{\sum n_i}
\end{aligned}
\end{equation}

\begin{table*}\label{tab:MFP}
\begin{tabular}{|c|c|c|c|c|c|c|c|c|}
    \colrule
    Type          & K  & $\alpha$   & $\beta$    & $\gamma$  &  $\mu_{1}$  & $\mu_{2}$    & $C_{1}$  & $C_{2}$\\
    &(MeV) & (GeV) & (GeV) & & (1/fm) & (1/fm) & (GeV) & (GeV)  \\
    \colrule
    MF             &270     &-0.209       &0.284      &7/6    &2.02 &1.0 &-0.383 &0.337                \\
   \colrule    
    \end{tabular}
\caption{Potential parameters in mean field mode.}
\end{table*}

\begin{table*}\label{tab:DATA_LIST}
\begin{tabular}{llll}
    \toprule
    Identifier                    &   Resonance weak decays   &  MB/MM scat.  &  Elastic scatting \\
    \colrule
    Full calc.                  &  Yes         &  Yes       & Yes            \\
    weak decays off       &  No           &  Yes       & Yes             \\
    MB/MM Scat. off        &  Yes         &  No        & Yes          \\
    Elas. Scat. off         & Yes           &   Yes      & No \\
    weak decays  \& Elas. Scat. off  & No & Yes & No \\
    \botrule
\end{tabular}
\caption{Simulation options used to study the effects of resonance weak decays and hadronic re-scattering. The data ``MB/MM Scat. off" means that we disable meson-baryon and meson-meson interactions, and only keep the baryon-baryon interactions. The ``elas. scat. off" is to disable hadronic elastic scattering and to keep only inelastic scattering. }
\end{table*}
The Delta theorem is usually used to evaluate statistical uncertainties of the cumulants and cumulants ratio~\cite{Luo:2011tp, Luo:2014rea}.

On the other hand, one can express the multi-particle correlation functions (also known as factorial cumulants) in terms of various order single particle cumulants (i.e.\ proton cumulants, but not net-proton cumulants)~\cite{Bzdak:2016sxg, Ling:2015yau, Kitazawa:2017ljq}
\begin{equation}\label{eq:cumulant_to_correlation}
\begin{aligned}
    \kappa_2 &= -\langle N\rangle + C_2 \\
    \kappa_3 &= 2\langle N\rangle - 3C_2 + C_3 \\
    \kappa_4 &= -6\langle N\rangle + 11C_2 - 6C_3 + C_4 
\end{aligned}
\end{equation}
Thus, we also have: \begin{equation}\label{eq:correlation_to_cumulant}
\begin{aligned}
    C_2 =& \langle N \rangle + \kappa_2  \\
    C_3 =& \langle N \rangle + 3\kappa_2 + \kappa_3 \\
    C_4 =& \langle N \rangle + 7\kappa_2 + 6\kappa_3 + \kappa_4
\end{aligned}
\end{equation}
, where the $\kappa_n$ are used to denote various order correlation functions (or factorial cumulants). The $\kappa_n$ ($n>2$) of Poisson distributions is always zero. Thus, one can measure non-Poisson fluctuations from correlation functions. The correlation functions can be calculated by factorial moments $F_n$ as: 
\begin{equation}\label{eq:factorial_moments}
    F_n = \langle N^n\rangle_\mathrm{f} \equiv \langle N(N-1)\cdots(N-n+1) \rangle
\end{equation}
The relations between factorial moments and correlation functions are equivalent to those between moments and cumulants. Comparing with Eq. (\ref{eq:moment_to_cumulant}), we have
\begin{equation}\label{eq:factorial_moment_to_correlation}
\begin{aligned}
    \kappa_2 &= F_2 - F_1^2 \\
    \kappa_3 &= 2F_1^3 - 3FF_2 + F_3
\end{aligned}
\end{equation}
It was predicted that the critical fluctuations can be encoded in the acceptance dependence of cumulants and/or correlation functions~\cite{Ling:2015yau, Bzdak:2016jxo}.  We found that that the enhancement of $\kappa\sigma^2$ at 7.7 GeV in most central Au+Au collisions observed by the STAR experiments is mainly due to the four-particle correlation function~\cite{Bzdak:2016sxg}. In our previous studied with UrQMD model~\cite{He:2017zpg}, we observed large deviations from experimental results in second and fourth order correlation functions. Thus, it is important to study the correlation functions to understand different non-critical contributions.

\begin{figure*}\label{fig:DNDY}
    \subfloat[]{\label{fig:EOS_DNDY}
        \includegraphics[width=0.42\textwidth]{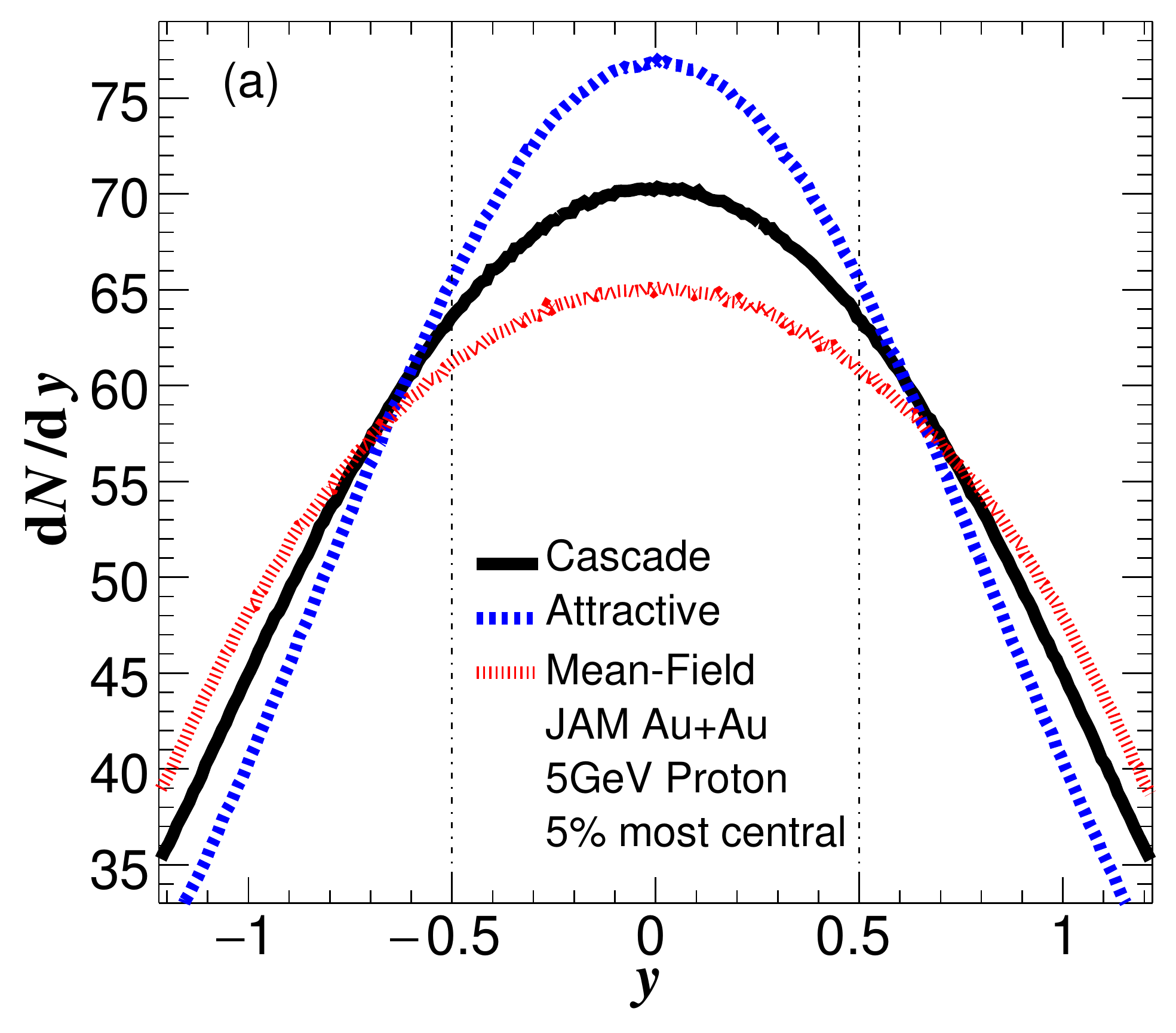}
    }
    \subfloat[]{\label{fig:SCAT_DNDY}
        \includegraphics[width=0.42\textwidth]{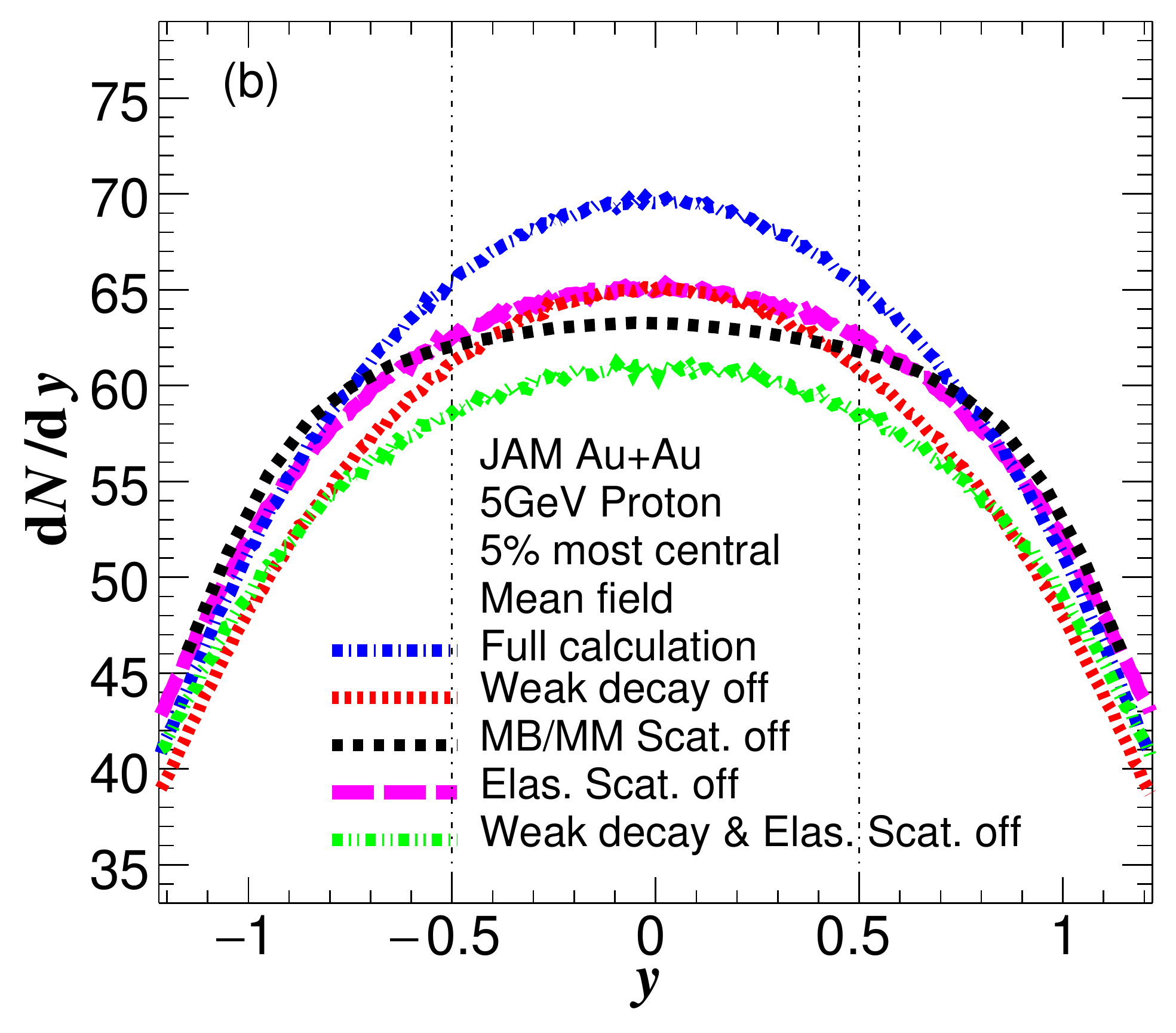}
    }
\caption{Rapidity ($\mathrm{d}N/\mathrm{d}y$) distributions for proton in most central (0-5\%) Au+Au collisions at $\sqrt{s_\mathrm{NN}} = 5$~GeV. (a) Different EoS implemented in JAM model (cascade, attractive re-scattering orbit, and mean-field). (b) $\mathrm{d}N/\mathrm{d}y$ distributions with/without weak decays, MB/MM scattering and elastic scattering.}
\end{figure*}

\begin{figure*}\label{fig:EVENT_BY_EVENT}
    \subfloat[]{\label{fig:EOS_E2N}
        \includegraphics[width=0.42\textwidth]{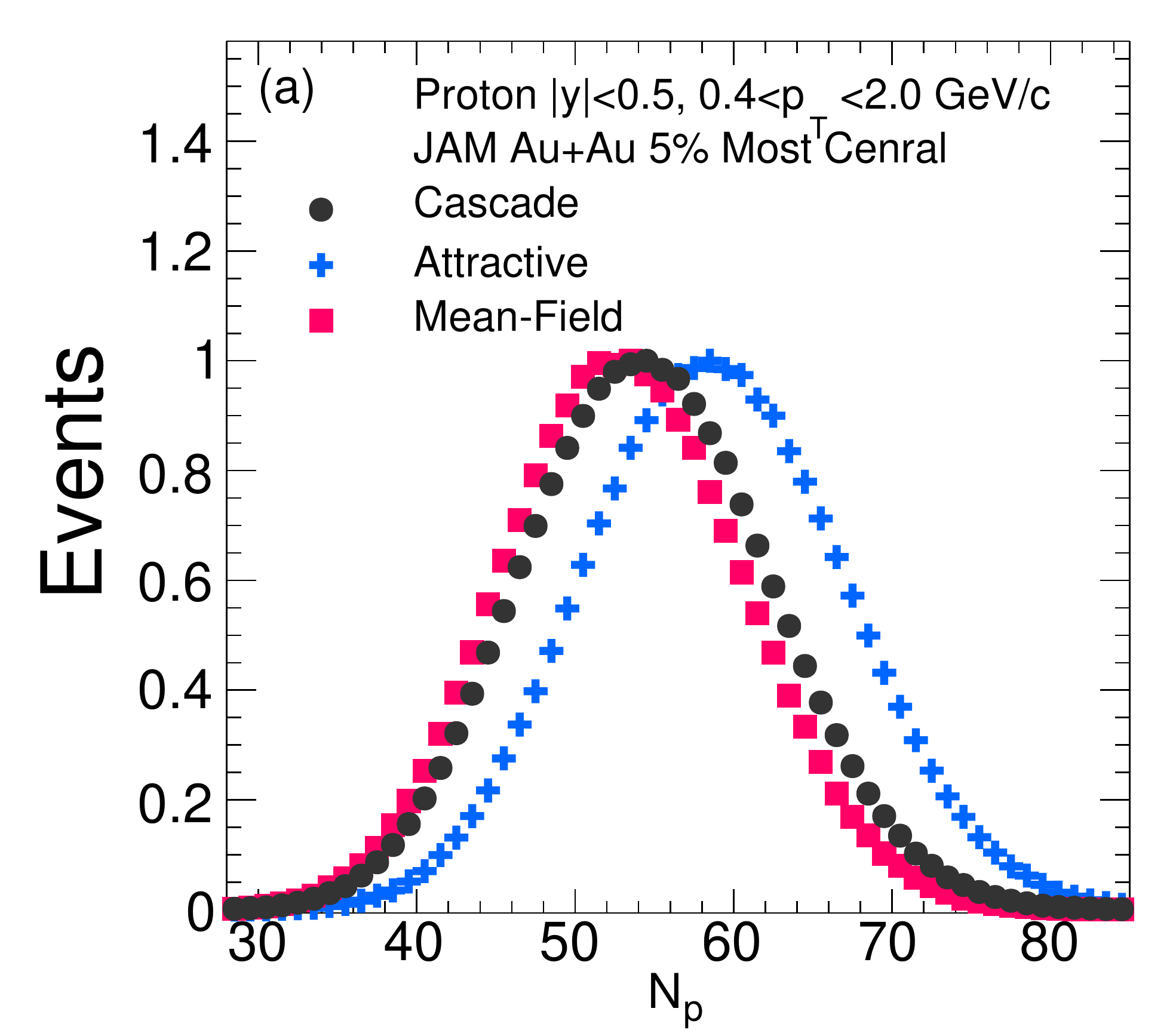}
    }
    \subfloat[]{\label{fig:SCAT_E2N}
        \includegraphics[width=0.42\textwidth]{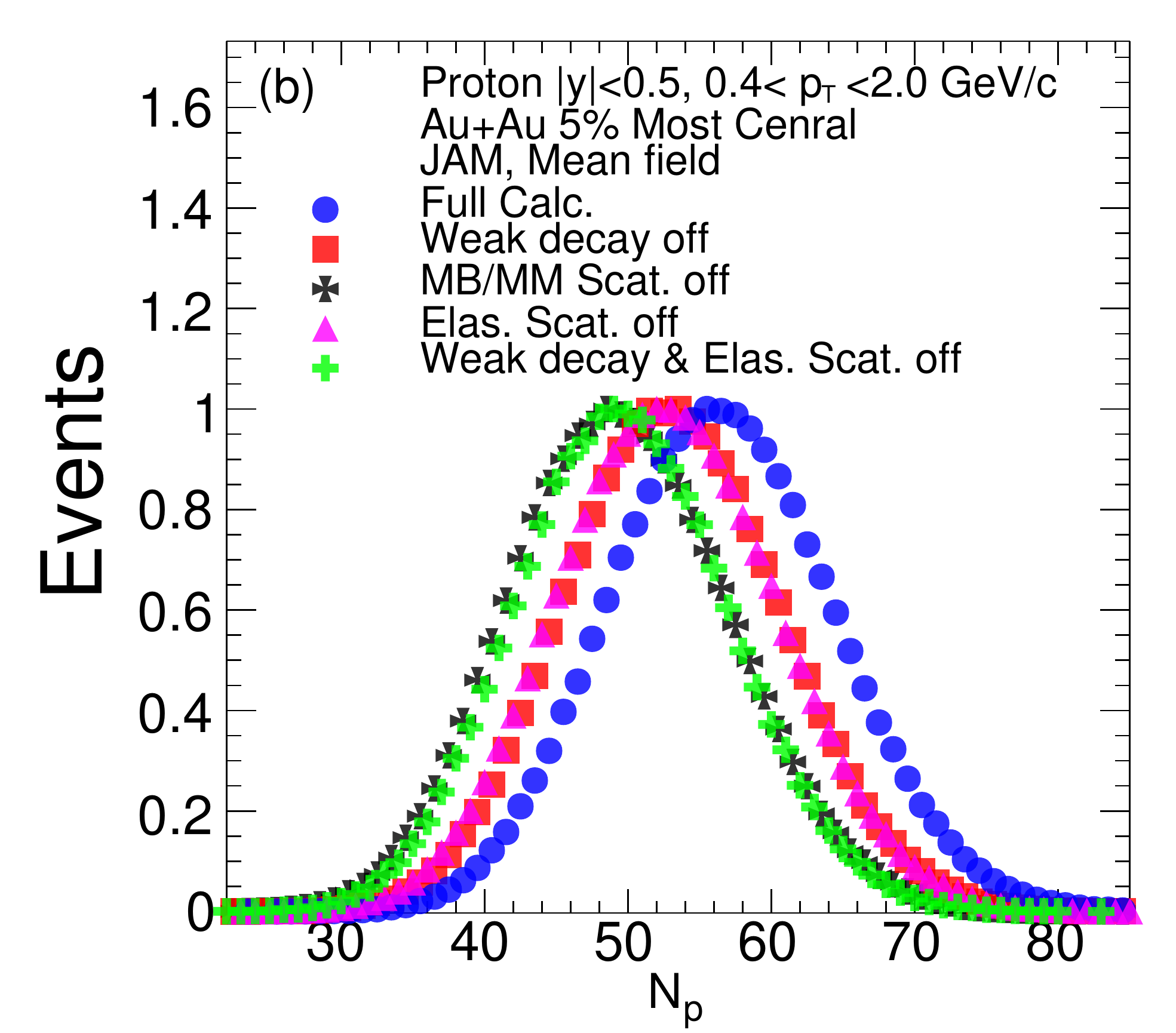}
    }
\caption{Normalized event-by-event proton multiplicity distributions in most central (0-5\%) Au+Au collisions at $\sqrt{s_\mathrm{NN}} = 5$~GeV. The $N_\mathrm{p}$ represents the proton number in an event. (a) Different EoS implemented in JAM model (cascade, attractive re-scattering orbit, and mean-field). (b) With/without weak decays, MB/MM scattering and elastic scattering.}
\end{figure*}
\section{JAM Model}\label{sec:effect_in_jam}
JAM (Jet AA Microscopic Transportation Model) is a simulation program which is designed to simulate relativistic nuclear collisions from initial stage of nuclear collision to final state interaction in hadronic gas state. In JAM mode,  hadrons and their excited states have explicit space and time evolution trajectories by the cascade method. Inelastic hadron-hadron collisions are modeled with resonance at low energy, string picture at intermediate energy and hard parton-parton scattering at high energy. In JAM model, the nuclear mean-field is implemented based on the simplified version of the relativistic quantum molecular dynamics (RQMD) approach. It is a skyrme-type density dependent and Lorentzian-type momentum dependent scalar mean-field potential~\cite{Isse:2005nk}. More features can be seen in references~\cite{Nara:1999dz, Nara:2016phs, Nara:2016hbg, Hirano:2012yy}. 
In JAM model,  one can study the effects of various type of equation of state (EoS). 

Generally, the EoS of medium can be expressed in the relation between pressure and the energy density of system: $p=p(\epsilon)$. The pressure of system can be given by virial theorem~\cite{Sorge:1998mk}
\begin{equation} \label{eq:virial_theorem}
    P = P_\mathrm{f} + \Delta P
 \end{equation}
where $P_\mathrm{f}$ is free stream part and the $\Delta P$ is determined by the momentum transfer in two-body collision.  The $\Delta P$ can be reduced by introducing an attractive scattering angle, while it is increased by selecting a repulsive scattering orbit. In JAM model, the attractive scattering orbit is used to simulate the effect of softening of EoS for the first-order phase transition. For a cascade mode, the azimuthal angle of the two-body collision is chosen randomly.  It means we select attractive or repulsive orbit is of equal chance, which lead to the free hadron gas EoS. In the mean-field mode, nucleons feel repulsive interactions with other particles. Therefore, the $\Delta P$ in Eq. (\ref{eq:virial_theorem}) is enhanced and we get a stiffer EoS. In this work, we use mean field mode with parameters shown in~\autoref{tab:MFP} . The results from attractive orbit and mean-field modes are compared with the results from default cascade mode, separately. To study the effects of resonance weak decays and hadronic re-scattering, we have produced five types of JAM model data with mean field EoS, which is shown in \autoref{tab:DATA_LIST}.

\section{Results}\label{sec:effect_to_fluctuation}
\subsection{Proton $dN/dy$ and Event-by-Event Distributions}
In this section, we will discuss the proton $\mathrm{d}N/\mathrm{d}y$ and the event-by-event proton number distributions in 0-5\% most central Au+Au collisions at  $\sqrt{s_\mathrm{NN}} = 5$ GeV from different cases. Figure~\ref{fig:EOS_DNDY} shows the proton $\mathrm{d}N/\mathrm{d}y$ distributions from three types of EoS. By comparing the distributions from different EoS, we found more protons are stopped at mid-rapidity due to the softening of EoS realized by using the attractive orbit scattering. However, due to the repulsive interactions, lower mean value of $\mathrm{d}N/\mathrm{d}y$ distribution is observed for the mean field mode.

\begin{figure*}\label{fig:CUMULANTS_AND_CORRELATION}
       \hspace*{1em}\includegraphics[width=0.985\textwidth]{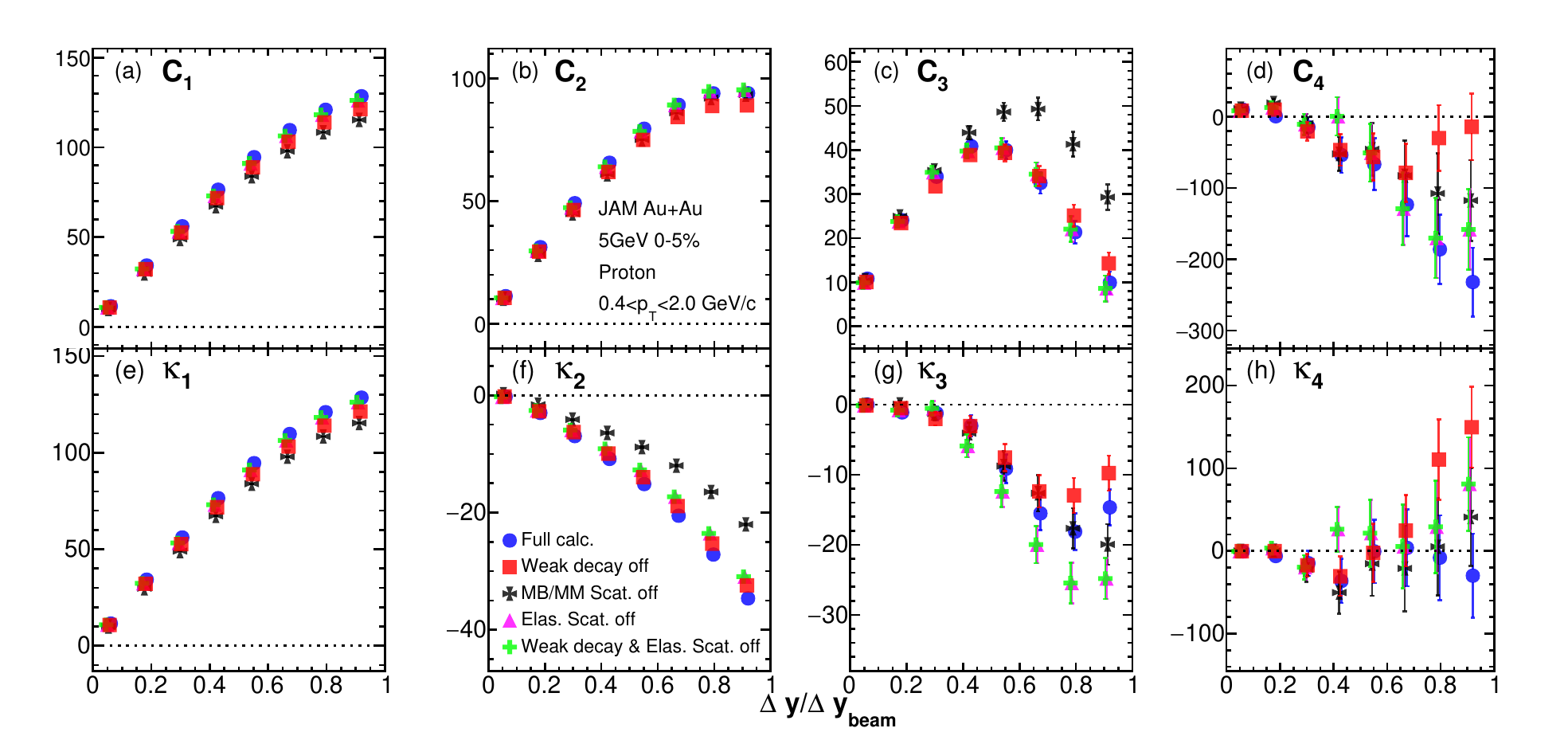}
\caption{Rapidity acceptance dependence of proton cumulants ($C_1$ $\sim$ $C_4$) and correlation functions ($\kappa_1$ $\sim$ $\kappa_4$) in 0-5\% most central Au+Au collisions $\sqrt{s_\mathrm{NN}} = 5$~GeV. The results are obtained with/without weak decays, MB/MM scattering and elastic scattering from JAM model. In X-axis label,  $\Delta y = 2y'$ denotes $|y|<y'$ in calculations, and the $y_\mathrm{beam}=1.63$ ($\Delta y_\mathrm{beam}$=3.26) is the beam rapidity for $\sqrt{s_\mathrm{NN}} = 5$ GeV.}
\end{figure*}

In Fig.~\ref{fig:SCAT_DNDY}, we compared the proton $\mathrm{d}N/\mathrm{d}y$ distributions from resonance weak decays and hadronic re-scattering.
It was observed that the proton $\mathrm{d}N/\mathrm{d}y$ distributions show a significant decrease when the weak decays switched off in JAM model. 
On the other hand, the effects of hadronic re-scattering are studied via disabling the meson-baryon (MB) and meson-meson (MM) interactions, and the elastic scattering among hadrons. We found the $\mathrm{d}N/\mathrm{d}y$ distributions from the two cases become flatter and wider than the distribution from the full calculation. This is due to the reduced baryon stopping caused by switching off the MB/MM interactions and/or elastic re-scattering. In the case of switching off the MB/MM interactions, only the baryon-baryon (BB) interactions and corresponding string excitation/fragmentation are playing a decisive role during the heavy-ion collision process. 

Before discussing the results of proton cumulants and correlation functions, we show the event-by-event proton number distributions for different cases in 0-5\% most central Au+Au collisions at  $\sqrt{s_\mathrm{NN}} = 5$ GeV. Figure~\ref{fig:EOS_E2N} shows event-by-event proton number distributions for different EoS. We observed that a softer EoS (attractive scattering orbit) tends to have more protons stopped at mid-rapidity and the proton number distribution has a larger mean value than the results from cascade mode, while a stiffer EoS (mean-field, or repulsive potential) leads to a smaller mean value~\cite{He:2016uei}. In Fig.~\ref{fig:SCAT_E2N}, it is shown that the effects of weak decays can enhance the proton multiplicities at mid-rapidity region similarly to switching on MB/MM scattering or elastic scattering. In Ref.~\cite{He:2016uei}, we concluded that the effects of mean field (only include scalar interactions) and attractive re-scattering orbit on proton number fluctuations are not significant and cannot lead to large proton $C_4$ or $\kappa \sigma^{2}$ at low energies. This might indicate the current JAM model doesn't capture the essential physics or true EoS that dominated the large increase in proton fourth order cumulant $C_4$.  For example, currently, only momentum dependence scalar potential is included in the mean field, but the vector potential could be also important. For future work, it would be interesting to study the mean field effects by including both the scalar and vector potential. More importantly, there is no physics of phase transition and critical point implemented in the JAM model.

In the following, we focus on discussing the effects of resonance weak decays and hadronic re-scattering on proton number fluctuations.

\subsection{Rapidity acceptance dependence of proton cumulants and correlation functions}
Theoretically, it was predicted that the rapidity acceptance dependence of the proton cumulants and correlation functions are important observables to search for the QCD critical point and understand the smearing/non-equilibrium effects of dynamical expansion on the fluctuations in heavy-ion collisions~\cite{Ling:2015yau,Ohnishi:2016bdf,Bzdak:2016sxg,Bzdak:2017ltv,Mukherjee:2015swa,Brewer:2018abr}. Due to the long range correlations near the critical point, it is expected that the proton cumulants ($C_n$) and/or multi-proton correlation functions ($\kappa_n$) will be dominated by critical behavior, which shows power law dependence as a function of number of protons and/or rapidity acceptance as $C_n,\kappa_n \propto (N_{p})^{n} \propto (\Delta y)^{n}$~\cite{Ling:2015yau}. This requires the typical correlation length of the system near the critical point is larger than the rapidity interval ($\Delta y < \xi $). If the rapidity acceptance is further enlarged and the $\Delta y$ becomes much larger than $\xi$ ($\Delta y \gg \xi$), the proton cumulants and/or multi-proton correlation functions will then be dominated by statistical fluctuations, which results in $C_n,\kappa_n \propto N_{p}^{} \propto \Delta y^{}$. However, the rapidity acceptance of the proton cumulants and multi-proton correlation functions are also sensitive to the background effects, such as baryon number conservation (BNC). Thus, by comparing the acceptance dependence of proton cumulants and multi-proton correlation from various simulation options from JAM model, we can clearly demonstrate effects of BNC and other background effects, such as the equation of states,  resonance weak decays and hadronic re-scattering effects. 
\begin{figure}\label{fig:CUMULANT_RATIO}
\includegraphics[width=0.5\textwidth]{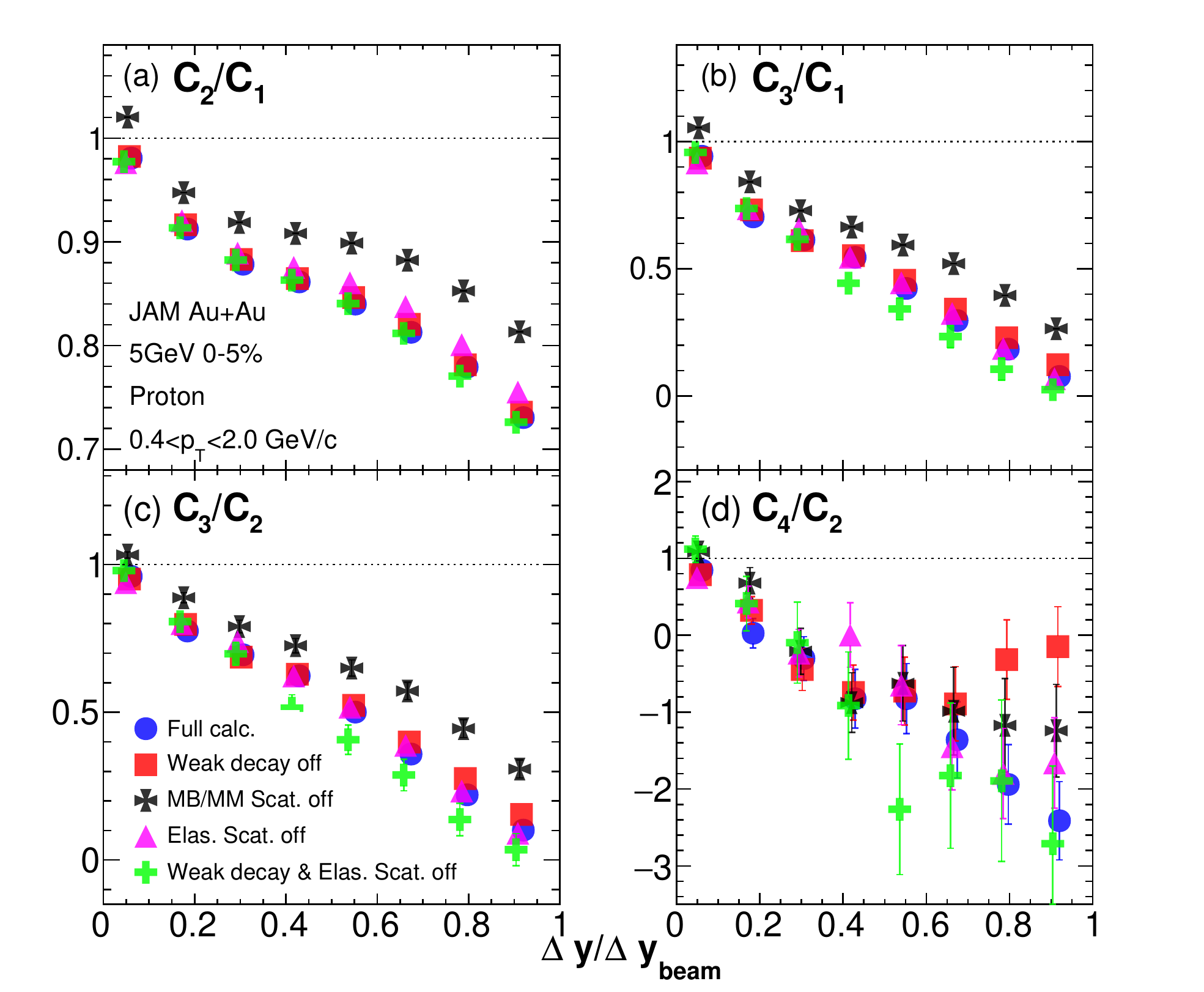}
\caption{Rapidity acceptance dependence of cumulant ratios of proton multiplicity distributions in the 0-5\% most central Au+Au collisions at $\sqrt{s_\mathrm{NN}} = 5$~GeV from JAM model.
To study the effects of weak decays and hadronic re-scattering, we compared the results from five different types of the data generated by JAM model. }
\end{figure}
\begin{figure*}\label{fig:energy_dep}
\includegraphics[width=0.6\textwidth]{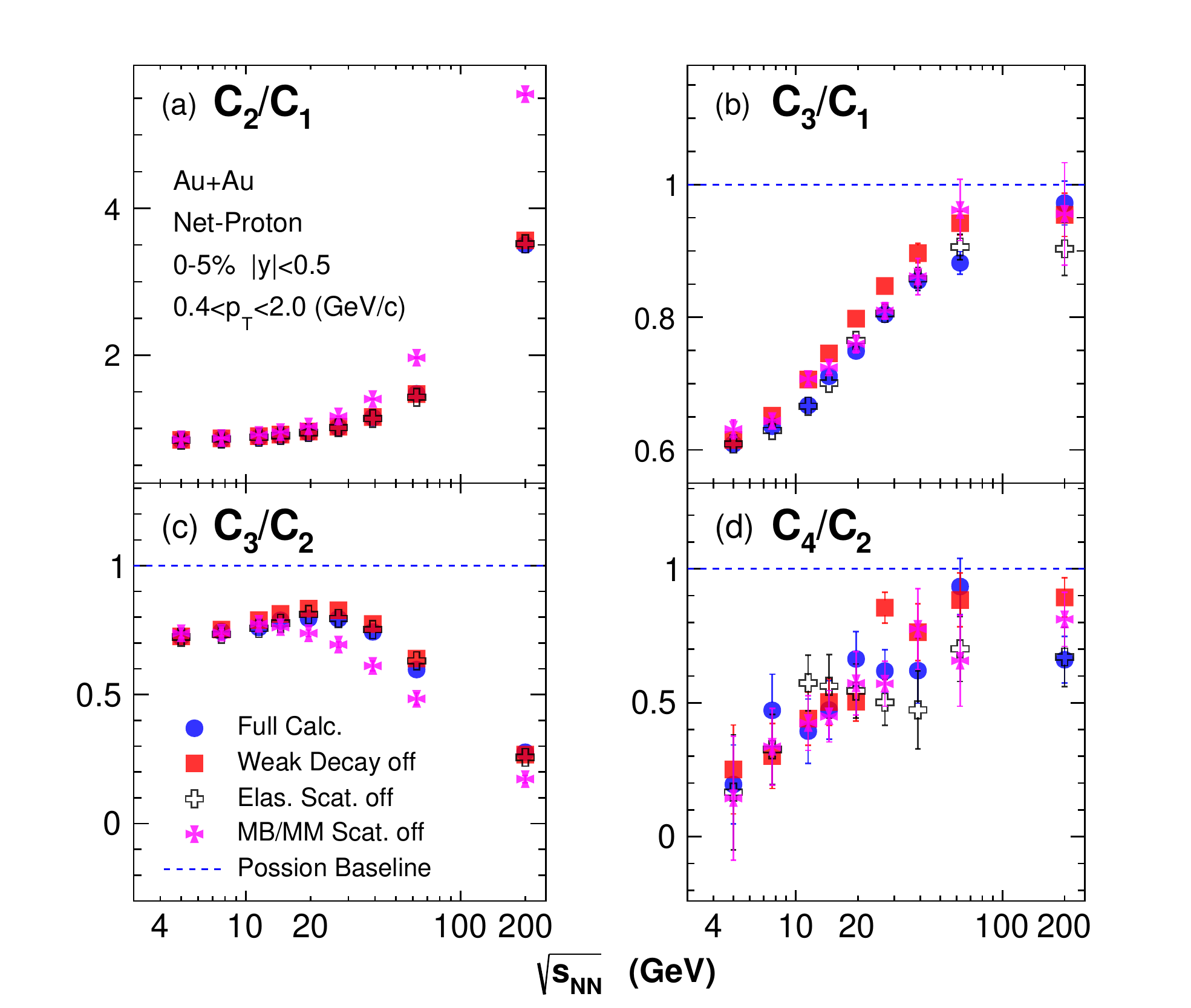}
\caption{Energy dependence of cumulants ratios of net-proton multiplicity distributions in 0-5\% most central Au+Au collisions at $\sqrt{s_\mathrm{NN}}$ = 5, 7.7, 11.5, 14.5, 19.6, 27, 39, 62.4, 200 GeV from JAM model.
To study the effects of weak decays and hadronic re-scattering, we compared the results from four different types of the data generated from JAM model. }
\end{figure*}
Figure~\ref{fig:CUMULANTS_AND_CORRELATION} shows rapidity acceptance dependence of various orders of proton cumulants and correlation functions in 0-5\% most central Au+Au collisions at $\sqrt{s_\mathrm{NN}} = 5$~GeV. We observed that the effects of resonance weak decays and hadronic re-scattering on proton cumulants and correlation functions are small at mid-rapidity ($\Delta y/\Delta y_{beam}<0.3$), but those effects get larger when further increasing the rapidity coverage.  
By making comparison between results from different cases, we found the $C_1$ and $C_2$ values from full calculations are larger than other cases. This is mainly due to the resonance weak decays effects, especially the feed down contributions of protons from $\Lambda$ and $\Sigma^{+}$. At forward rapidity region, $\Delta y/\Delta y_{beam}>0.4$, the MB/MM scattering substantially suppresses the $C_3$ values, while resonance weak decays and elastic scattering have very small effects on $C_3$. For $C_4$, the results of different cases are consistent within statistical uncertainty.  In addition, due to the baryon number conservation, the third and fourth order proton cumulants show strong suppression when increasing the rapidity acceptance, as the effect of BNC becomes stronger when the fraction of proton number over total baryon in the acceptance gets larger~\cite{Bzdak:2012an}. 
In the second row of Fig.~\ref{fig:CUMULANTS_AND_CORRELATION}, we show various order proton correlation functions ($\kappa_1$ to $\kappa_4$) in 0-5\% most central Au+Au collisions at $\sqrt{s_\mathrm{NN}} = 5$~GeV from different cases. At mid-rapidity, we find that the effects of resonance weak decays and hadronic elastic scattering on the various order proton correlation functions are small. The resonance weak decays and hadronic re-scattering slightly suppress the two proton correlation function $\kappa_2$.  Due to BNC, the $\kappa_2$ is negative and monotonically decreases when enlarging the rapidity acceptance. This is because the BNC leads to anti-correlation of protons separated by any rapidity intervals. However, the $\kappa_3$ and $\kappa_4$ are almost flat and close to zero at mid-rapidity ($\Delta y/\Delta y_{beam}<0.3$), which means that the higher order ($n>2$) correlation functions are less sensitive to the effect of BNC. Furthermore, the MB/MM scattering  leads to larger suppression of the two particle correlation functions $\kappa_2$ than the results from the case by turning them off.  

In Fig.~\ref{fig:CUMULANT_RATIO}, we show the rapidity acceptance dependence of various order proton cumulant ratios in 0-5\% most central Au+Au collisions at $\sqrt{s_\mathrm{NN}}$ = 5 GeV. Generally, the proton cumulant ratios decrease when increasing the rapidity acceptance. This can be explained by the effects of BNC. The hadronic re-scattering via turning on MB/MM scattering suppresses the second and third order cumulant ratios ($C_2/C_1$,  $C_3/C_1$ and $C_3/C_2$). However, the resonance weak decays and elastic scattering have little influence on these cumulant ratios.

\subsection{Energy dependence of net-proton cumulant ratios}
As shown Fig.~\ref{fig:KV}, non-monotonic energy dependence of fourth order fluctuations $\kappa \sigma^{2}$ of net-proton multiplicity distribution is observed in the RHIC beam energy scan program. This observation is consistent with the theoretical expectations by assuming the presence of QCD critical point. However, one needs to study the background contributions to the observable carefully, especially to understand how those backgrounds depend on the collision energies. In this paper, we focus on discussing the effects of resonance weak decay and 
hadronic re-scattering. Fig.~\ref{fig:energy_dep} shows energy dependence of net-proton cumulant ratios in 0-5\% most central Au+Au collisions with four different cases. We found that the effects of hadronic elastic scattering on various order net-proton cumulant ratios are not significant. However, by switching off the MB/MM collisions, we observed the cumulant ratios $C_3/C_2$ are suppressed 
while the cumulant ratios $C_2/C_1$ are significantly enhanced, especially at high energy. On the other hand, the effects of resonance weak decays suppress the third order cumulant ratios ($C_3/C_1$ and $C_3/C_2$).  For $C_4/C_2$, it shows monotonic decreasing trend when decreasing the collision energy and the results from four different cases are consistent within statistical uncertainties. The  $C_4/C_2$ values are below Poisson baseline (unity) and cannot describe the non-monotonic energy dependence trend of $\kappa \sigma^{2}$ in most central Au+Au collisions observed in STAR data. 

As discussed in Ref.~\cite{He:2016uei}, the JAM model used in this analysis only includes the momentum dependence scalar potential in the mean field and doesn't implement the physics of critical point and phase transition. It would be interesting to study those effects on the proton number fluctuations in the future.

\section{Summary}
We studied the effects of resonance weak decays and hadronic re-scattering on proton cumulants and correlation functions in Au+Au collisions at $\sqrt{s_\mathrm{NN}}=5$ GeV within JAM model.  
For the hadronic re-scattering, we further studied the effects of MB/MM interactions and elastic hadronic scattering. In general, at mid-rapidity region, the effects of resonance weak decays and hadronic re-scattering on proton cumulants and correlation functions are small, but those effects get larger when further increasing the rapidity acceptance. The weak decays and hadronic re-scattering enhance the mean values and width of the proton number distributions at mid-rapidity Au+Au collisions, while those two effects slightly suppress the two particle correlation functions of protons. The MB/MM scattering suppresses the second and third order cumulant ratios ($C_2/C_1$,  $C_3/C_1$ and $C_3/C_2$). On the other hand, the baryon number conservation is a dominant background effect on the rapidity acceptance dependence of proton number fluctuations. It leads to a strong suppression of cumulants and cumulant ratios, as well as the negative proton correlation functions. However, the higher order correlation functions are less sensitive to the BNC. 
We also discussed the energy dependence of various order net-proton cumulant ratios in 0-5\% most central Au+Au collisions at $\sqrt{s_\mathrm{NN}}=5-200$ GeV.  We found that the effects of hadronic elastic scattering on various order net-proton cumulant ratios are not significant within statistical uncertainties and the resonance weak decays suppress the third order cumulant ratios ($C_3/C_1$ and $C_3/C_2$). The effects of switching off the MB/MM interactions significantly suppress the values of $C_3/C_2$ and enhance the values of $C_2/C_1$, especially at high energy. 
Due to the effect of BNC, the values of $\kappa \sigma^{2}$ ($C_4/C_2$) are significantly below Poisson baseline (unity) at low energies and cannot describe the non-monotonic energy dependence trend in most central Au+Au collisions observed in STAR data. Our work provides useful non-critical baselines for the future QCD critical point search in heavy-ion collisions at high baryon density region. 

\section{Acknowledgement}
This work is supported by the National Key Research and Development Program of China (2018YFE0205201),  the National Natural Science Foundation of China (No.11828501, 11575069, 11890711 and 11861131009). 

\bibliography{ref}

\end{document}